\documentclass[10pt,a4paper,twoside]{article}
\usepackage{epsfig}

\begin{document}

{\LARGE \bf Measurements of strong magnetic fields in umbra of
sunspots: Crimea vs Mt. Wilson}

\bigskip

Yu.T. Tsap$^{1,2}$, V.A.~Perebeynos$^{1}$,
A.V.~Borisenko$^{1}$, N.I. Lozitska$^3$, N.I.~Shtertser$^1$, G.G. Motorina$^{2,4}$, A.I.~Kuleshova$^2$\\

$^1$Crimean Astrophysical Observatory (Russian Academy of
Sciences), Crimea, Russia, yur@craocrimea.ru

$^2$Central Astronomical Observatory at Pulkovo of Russian Academy of Sciences, St. Petersburg, 196140, Russia

$^3$Astronomical Observatory of Taras Shevchenko National
University of Kyiv, Kyiv, Ukraine

$^4$Astronomical Institute, Academy of Sciences of the Czech Republic, 251 65 Ondrejov, Czech Republic

\bigskip

{\bf Abstract.}

The comparative analysis for 1324 measurements
 of the corresponding sunspot magnetic fields with $B > 2.5$ kG (according to Crimean data) obtained
 at Crimean and Mt. Wilson observatories from 2010 to 2017 has been carried out.
It has been shown that the difference between measurements can
exceed 1 kG in some cases. The averaged values of the magnetic
field are equal to 2759~G (Crimea) and 2196~G (Mt. Wilson).
The maximum sunspot magnetic field measured at Mt. Wilson does
not reach 2.7~kG while according to Crimean data it can exceed
4.0~kG. The correlation coefficient between measurements of
magnetic fields in different observatories does not exceed 0.22.
The probable reasons of significant discrepancies are discussed.

\bigskip

Key words: Sun: sunspot magnetic fields

\bigskip

{\bf Introduction}

\medskip

Sunspots are the largest magnetic flux concentrations in the solar
photosphere. The magnetic energy density inside sunspots is higher
than the kinetic energy density, resulting in a partial
suppression of the convection. Larger umbrae are darker and show a
higher magnetic field strength.

Visual (photographic) measurements of sunspot magnetic fields
using the Zeeman effect have continued until the present time at
Crimean Astrophysical Observatory (CrAO) and Mount Wilson
Observatory (MWO). Maximum distance between sigma components of
simple triplets is measured in the profile of the line. Then the measured size is transferred in
intensity of a magnetic field.

The longest series of observations of sunspot magnetic fields
exists at MWO (Hale et al., 1919; Livingston, 2006). The MWO
archive of drawings of umbrae, penumbra, and other magnetic
structures began in 1917. A majority of these drawings has been
digitized (scanned), and their images are available online
($ftp://howard.astro.ucla.edu/pub/obs/drawings$). In a typical
year, there are about 330 clear days at MWO.

Visual measurements have been carried out from 1955 at CrAO
(Severny, Stepanov, 1956; Stepanov, Petrova, 1958). The
observations have been carried out in the line FeI 6302 \AA\ since
1957. The data are presented in a kind of sketch of
the solar disk with all sunspot groups and their temporary
numbers. The drawings have been digitized (scanned) for data
beginning from 1984, and their images are available online
$http://solar.craocrimea.ru/eng/sunspots\_mf.htm\#archive$.
There are about 220 clear days at CrAO in a typical year.

Lozitska et al. (2015) compared measurements of sunspot magnetic
fields during 2010-2012. However, sunspots with strong and weak
magnetic fields where not separated. Besides, Lozitska et
al.(2015) compared the full number of measurements of the sunspot
magnetic field strengths at Mt. Wilson and Crimea but not
measurements of corresponding sunspots.

The aim of this paper is to provide a comparison between the
visual measurements at MWO and CrAO for corresponding sunspots
with strong ($> 2.5$ kG according to CrAO measurements) magnetic
fields.

\bigskip

{\bf Observations and data processing}

\medskip

\begin{figure}[ht]
\begin{center}
\includegraphics*[height=5.5 cm]{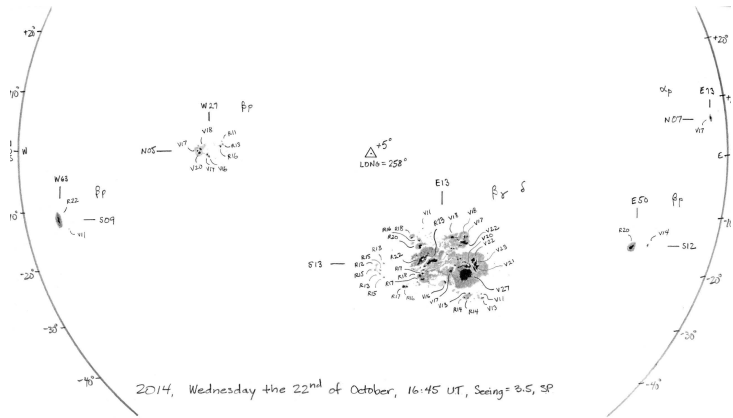}
\hspace{0.8cm}
\end{center}
\begin{center}
\caption{A sample daily drawing of sunspots measured at MWO on
October 22, 2014. On the drawing West is at the left.
\label{Fig_1}
}
\end{center}
\end{figure}

The daily observations at MWO are performed at the 150-foot (45.7
m) Solar Tower (ST, $http://obs.astro.ucla.edu/150_tele.html$).
Solar images are constructed using a coelostat and a lens, and
have a diameter of about 42 cm. When taking a measurement, the
observer marks the boundary of the solar disk, and draws the
position and configuration of sunspots. Magnetic field
measurements are then carried out by measuring the splitting of
the Zeeman components. The intensity of the magnetic field at the
center of the sunspot was measured visually using the iron
spectral line FeI 5250 \AA\ line with Lande factor 3.0. An image
of the sunspot to be measured passes through a polarization
analyzer placed in front of the slit of the 75-foot (22.9 m)
spectrograph. The analyzer alternately transmits left and right
circularly polarized light in narrow strips along the slit. The
micrometer is placed at the focus of the spectrograph and, by tilt
of the glass plate, shifts the wavelength position of the image of
one strip until the two oppositely polarized Zeeman sigma
components in adjacent strips coincide. The amount of shift
corresponds to the field strength. With this setup, sunspot field
strengths can be obtained at the accuracy of hundreds gauss. The
measured value of the magnetic field is then denoted on the
drawing of the respective sunspot umbra (Fig.\ref{Fig_1}).

\begin{figure}[ht]
\begin{center}
\includegraphics*[height=8 cm]{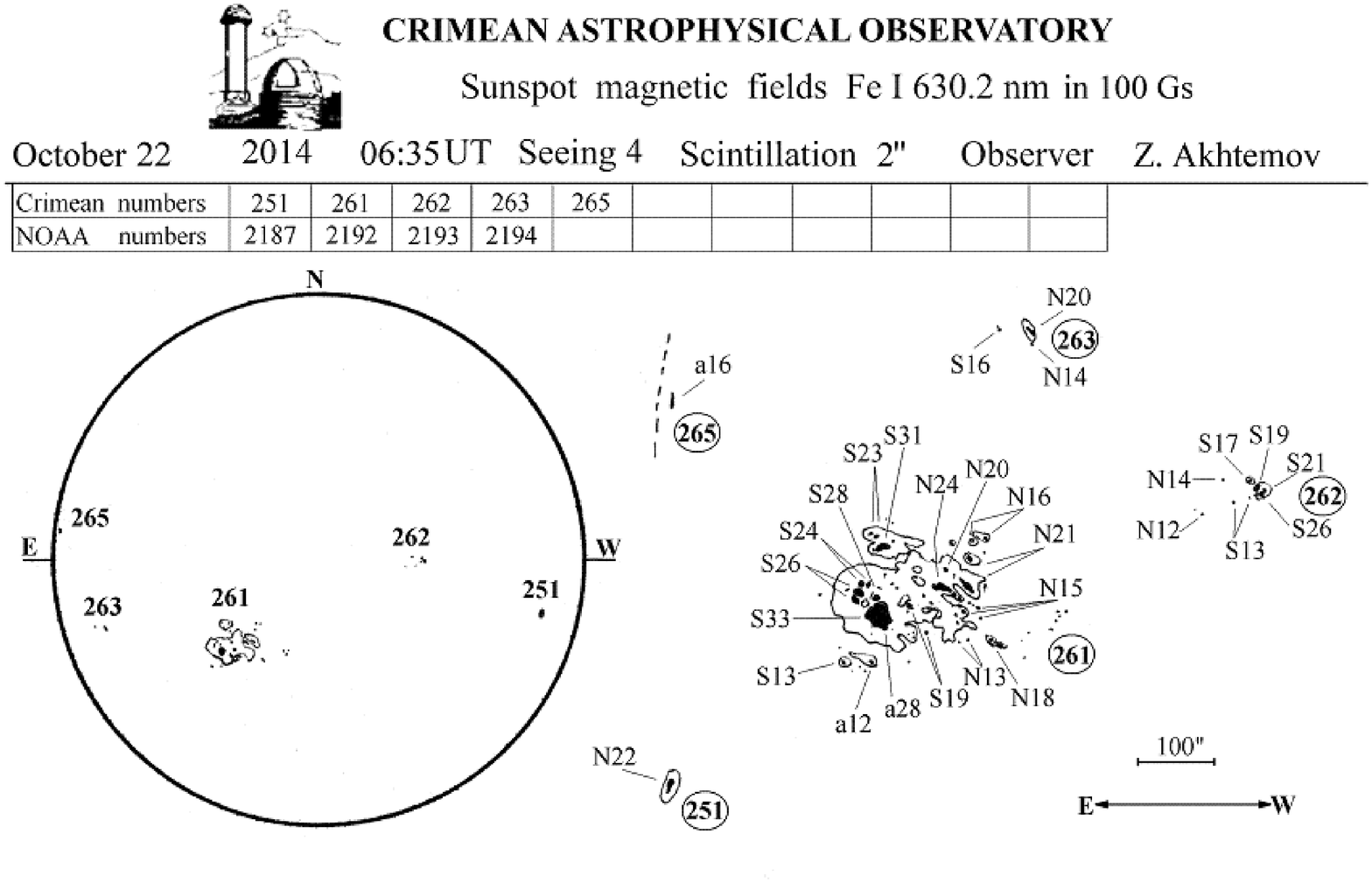}
\hspace{0.8cm}
\end{center}
\begin{center}
\caption{A sample daily drawing of sunspots measured at CrAO on
October 22, 2014.
\label{Fig_2}
}
\end{center}
\end{figure}

Visual measurements of sunspot magnetic fields are carrying out at
Tower Solar Telescope (TST-2) in the CrAO from 1955. The 60-cm
coelostat, 45-cm spherical primary mirror (f/27), one flat and two
convex secondary mirrors provide f/27, f/46 and f/78 Cassegrain
foci (12, 21 and 30 m) on the entrance slit of a spectrograph.
There are an echelle-grating, Universal Spectrophotometer with a
scanning system and a CCD camera. Solar images have diameters
from 8 to 30 cm. Observations have been carried out in the spectral
line FeI 6302 \AA\ with Lande factor 2.5 since 1957. Then the
measured distance is transferred in intensity of a magnetic field
using the special table and then denoted on the drawing of the
respective sunspot umbra. The procedure of daily observations at
TST-2 and ST (MWO) is similar. The sunspot field strengths can
be obtained at the accuracy of one hundred gauss. Note that the
sunspot penumbras as distinguished from MWO are not drawn at
CrAO. Besides, the daily drawings consist of two pictures of
different scales (Fig.\ref{Fig_2}).

\begin{figure}[ht]
 \centerline{\hspace*{-0.02\textwidth}
        \includegraphics[width=0.70\textwidth,clip=]{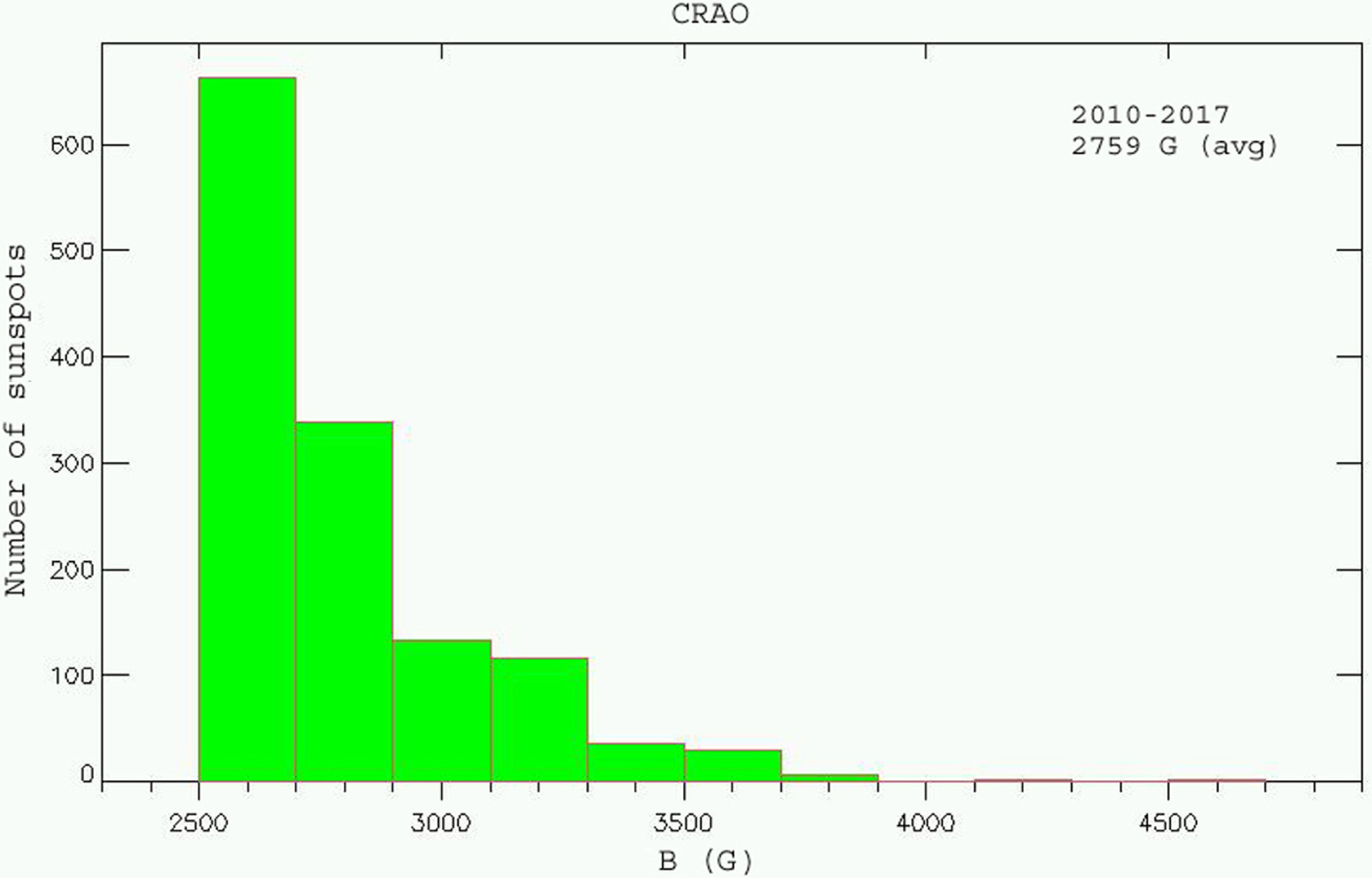}}

      \centerline{\hspace*{-0.02\textwidth}
       \includegraphics[width=0.70\textwidth,clip=]{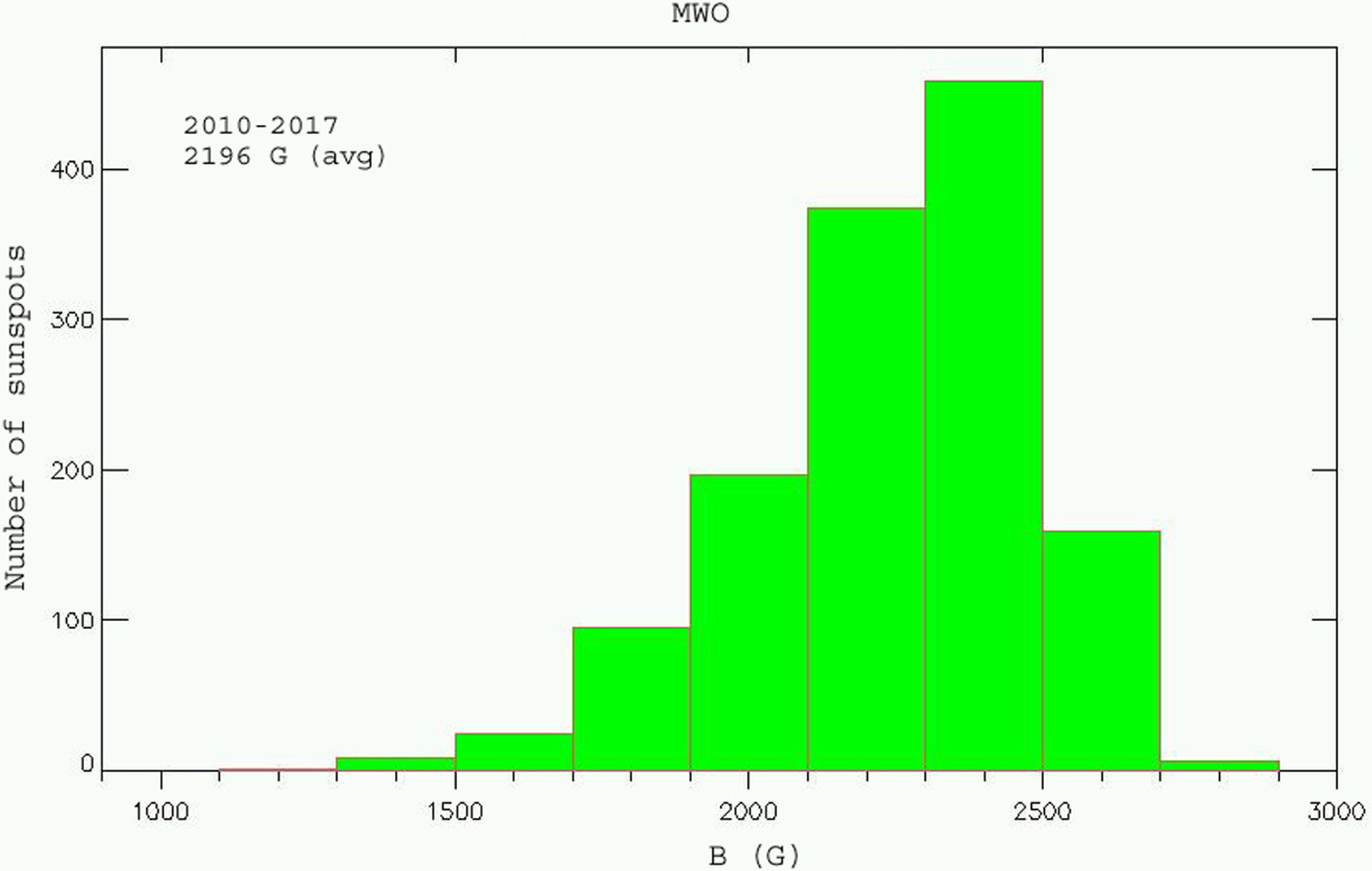}}
\hspace{0.8cm} \caption{Comparison of corresponding sunspot
magnetic field measurements at CrAO and MWO in eight intervals of
the magnetic field.
\label{Fig_3}
}
\end{figure}

In order to compare the sunspot magnetic fields from CrAO and MWO
we selected magnetic field values greater than 2.5 kG according to
CrAO measurements from 31 July 2010 to 01 October 2017. After that
we selected corresponding sunspot magnetic fields measured in MWO.
As result, we found 1324 corresponding sunspots (see Tab.1 at
$http://solar.craocrimea.ru/eng/observations.\\htm$). The typical
time difference between measurements was about 10 hours. It should
be emphasize that according to the CrAO data we found about 134
sunspots with the magnetic fields greater than 3 kG. The most
strongest magnetic fields were measured on 22 May 2016 in NOAA
12546 (4.1 kG) and on 3 September 2017 in NOAA 12674 (4.6 kG).
In turn, the magnetic fields measured at MWO do not exceed 2.7
kG. The diagrams of magnetic field measurements for corresponding
sunspots are presented in Fig.\ref{Fig_3}. The average value of CrAO
measurements is 2759 G, while the average value of MWO
measurements is 2196 G. Thus, the difference between measurements
exceeds 500 G. Besides, the correlation coefficient between MWO
and CrAO measurements turned out to be about 0.22. These results suggest that MWO data can not be used for sunspots with strong magnetic fields.

\bigskip

{\bf Discussion and conclusions}

\medskip

In spite of the significant progress in the sunspot magnetic field
measurements, the visual method based on the measurements of the
Zeeman splitting gives the most reliable results. In fact, visual
measurements of magnetic field strengths in sunspot umbra provide
data on magnetic field strength modulus directly, i.e.,
irrespective from any solar atmosphere model assumption. In
addition, results of measurements are not affected by the signal
saturation for strong magnetic fields, low light level,
instrumental polarization etc. Unlike magnetographic measurements,
these data do not need various calibration curves for different
sunspots and other features of the Sun.

Taking into account the above-mentioned and the similarity of
methods of measurements we have compared results of strong
magnetic fields of sunspots measured at CrAO and MWO. Although
the difference between the formation heights of lines 6302 \AA\
and 5250 \AA\ is less than 30 km and the time of measurements of
corresponding sunspots does not exceed 15 hours, the results of
measurements of strong magnetic fields at CrAO and MWO turned
out to be quite different.

It seems to us, this discrepancy can be caused by the
 small thickness of tip plate used at MWO, and, because of that, the visual measurements of
 strong magnetic fields becomes impossible (A.Pevtsov, private communication).
Also some problems can be caused by the calibration of
 measurements at CrAO and MWO (Livingston et al., 2006; Lozitska et al., 2015). We hope to
 consider these questions in future.

\bigskip

{\bf Acknowledgements}

We acknowledge A.N. Babin, A.N. Koval, and A. Pevtsov for
fruitful discussions. Research was partially supported by the
Fundamental Research Program of Presidium of the RAS N 28. GM was supported by the project RVO:67985815.

\newpage

{\bf References}

\medskip

Hale, G.E., Ellerman, F., Nicholson, S.B., Joy, A.H., 1919,
Astrophys. J., 49, 153.

Livingston, W., Harvey, J.W., Malanushenko, O.V., Webster, L.,
2006, Solar Phys. 239, 41.

Lozitska, N.I., Lozitsky, V.G., Andryeyeva, O.A., Akhtemov, Z.S.,
Malashchuk, V.M., Perebeynos, V.A., Stepanyan, N.N., Shtertser,
N.I., 2015, Adv. Space Res., 55, 897.

Severny, A.B., Stepanov, V.E., 1956, Izv. Krim. Astrofiz. Obs.,
16, 3.

Stepanov, V.E., Petrova, N.N., 1958, Izv. Krim. Astrofiz. Obs.,
18, 66.

\end{document}